\documentclass[journal]{IEEEtran}
\usepackage{cite}
\usepackage{amsmath,amssymb,amsfonts}
\usepackage{algorithmic}
\usepackage{graphicx}
\usepackage{float}
\usepackage{textcomp}
\usepackage{xcolor}
\usepackage{subfigure}
\usepackage{caption}
\usepackage{url}
\usepackage{algorithm}
\usepackage{amsmath}
\usepackage{url}
\usepackage{booktabs}%修改表格线段的粗细
\usepackage{hyperref}%自动索引
\usepackage[utf8]{inputenc}
\def\BibTeX{{\rm B\kern-.05em{\sc i\kern-.025em b}\kern-.08em
    T\kern-.1667em\lower.7ex\hbox{E}\kern-.125emX}}
\begin{document}

\title{UAV-aided Metaverse over Wireless Communications: A Reinforcement Learning Approach}
\author{\IEEEauthorblockN{Peiyuan Si$^1$,
Wenhan Yu$^1$,
Jun Zhao$^1$,
Kwok-Yan Lam$^1$,
Qing Yang$^2$}
\IEEEauthorblockA{\\$^1$School of Computer Science \& Engineering\\Nanyang Technological University, Singapore\\
$^2$University of North Texas, United States\\
\{peiyuan001, wenhan002\}@e.ntu.edu.sg, \{junzhao, kwokyan.lam\}@ntu.edu.sg}, Qing.yang@unt.edu}

\maketitle
\begin{abstract}
Metaverse is expected to create a virtual world closely connected with reality to provide users with immersive experience with the support of 5G high data rate communication technique. A huge amount of data in physical world needs to be synchronized to the virtual world to provide immersive experience for users, and there will be higher requirements on coverage to include more users into Metaverse. However, 5G signal suffers severe attenuation, which makes it more expensive to maintain the same coverage. Unmanned aerial vehicle (UAV) is a promising candidate technique for future implementation of Metaverse as a low-cost and high-mobility platform for communication devices. In this paper, we propose a proximal policy optimization (PPO) based double-agent cooperative reinforcement learning method for channel allocation and trajectory control of UAV to collect and synchronize data from the physical world to the virtual world, and expand the coverage of Metaverse services economically. Simulation results show that our proposed method is able to achieve better performance compared to the benchmark approaches.
\end{abstract}

\begin{IEEEkeywords}
Metaverse, UAV, cooperative reinforcement learning, PPO
\end{IEEEkeywords}

\section{Introduction}

%metaverse
The proposal of Metaverse has been promoted by the implementation of 5G communication technology and maturing AR/VR devices in recent years \cite{wang2022mobile, AllYouNeedToKnow, GlobeCom2022, Terence2022resource}. Metaverse aims to create a virtual world for all kinds of activities, including education, trading and gaming, and is considered the next generation of the Internet \cite{MetaEdu2020, MetaEdu2022, MetaTrad_1, MetaGame}. With the support of AR/VR applications, online users are provided with immersive services that are similar to in-person activities, and the trading of virtual items brings job opportunities.

%data collection, coverage
To support the Metaverse applications, data synchronization and wide wireless network coverage are two practical problems to be solved as the Metaverse services usually involve wearable wireless devices. For the first problem, 5G communication technology is able to provide high-speed and low-latency data transmission, but it is not necessary to update all the collected data immediately, e.g., environment information to build the background of Metaverse and offline trading records \cite{5GSurvey, NFC, OfflineTrade}. For the second problem, 5G network suffers higher costs for the same coverage area due to severe signal attenuation. Thus, it is not economically efficient to deploy base stations in suburban with low population density, and in wild areas it is not even applicable to traditional base stations \cite{sensorWild}.

%uav
Unmanned aerial vehicle (UAV) is a cheaper substitution solution to set up network coverage for Metaverse data synchronization in the suburban area due to its ability to carry communication devices. The UAV technique has been fully studied and commercialized, and there are numerous works on UAV-based communication scenarios for traditional applications, e.g., research on communication resource allocation, UAV trajectory control and the internet of vehicles \cite{RuiZhangUAV, SiIoTJ, UAVIoV}.
%convex
The UAV-based optimization problems which take the trajectory of UAV into consideration usually segment the flight time of UAV into discrete time slots for the convenience of computation. The resource allocation variables need to be optimized in each time slot to obtain the global or local optimal. Although these methods ensure the convergence of the solution, the increasing number of time slots results to the increment of algorithm complexity. Besides, the integer characteristic of channel allocation variables results to mixed integer programming problems, which can be hard to solve if the variables are inseparable.

%RL
\textbf{Related Work.}
In some cases, reinforcement learning (RL) is more suitable for UAV-based optimization problems than convex methods because it gives a feasible solution with relatively good performance even if the global optimal is extremely hard to find, and it can handle time-sequential problems without increasing the number of variables.
Cui et al. \cite{Cui_MultiUAV} proposed multi-agent reinforcement learning resource allocation algorithm for multi-UAV networks, and showed fast convergence with the basic Q-learning algorithm.
Luong et al. \cite{luong2021deep} utilized the deep Q-learning algorithm to learn the network state for the decision of the movement of UAV, and improved the network performance by up to 70\%.
Rodriguez-Ramos et al. \cite{UAVland} implemented a versatile Gazebo-based reinforcement learning framework for UAV landing on a moving platform, which is a novel experiment of DDPG on UAV controlling research.

For communication optimization problems with discrete channels and continuous resource allocation, both discrete and continuous action spaces need to be considered. To solve discrete-continuous hybrid action space reinforcement learning problems, multi-agent architecture is commonly adopted. Fu et al. \cite{Hybrid_MultiAgent_1} proposed two multi-agent reinforcement learning architectures for hybrid action spaces based on deep Q-learning (DQN), where agents work in a parallel manner to generate joint actions. Jiang et al. \cite{Hybrid_multiAgent_2} designed a hybrid action algorithm for massive access control, which optimized the discrete action selection for back-off and distributed queuing problems and generate continuous action for access class barring.

%contribution
The agents of most existing hybrid action space reinforcement learning algorithms work in a parallel manner, which does not build the inter-agent relationship. In this paper, we propose a hybrid reinforcement learning architecture to optimize the discrete channel allocation variable and the continuous trajectory controlling variable. Two agents work in a sequential manner motivated by the alternative optimization algorithms, i.e., the output of an agent is the input of another agent. Compared to the existing works, our paper considers the inter-agent relationship for better convergence performance. The advantage of our scenario over traditional convex optimization is that the number of variables does not increase when the number of time slots increases, which is more friendly to time-sequential problems.

\textbf{Contribution.} The contributions of this paper are as follows:
\begin{itemize}
\item A PPO-based double-agent cooperative hybrid action reinforcement learning architecture (PPO-PPO) for UAV-enabled Metaverse data synchronization is proposed.
\item Proximal policy optimization (PPO) algorithm is implemented in both discrete action agents and continuous action agents, and two agents work in a sequential manner.
\item The simulation shows the comparison between the proposed algorithm and two baselines (DQN and duelling DQN), which verifies the advantage of our proposed PPO-PPO algorithm.
\end{itemize}

The rest of this paper is organized as follows. Section II introduces the proposed system model. The double-agent policy generation model and its implementation are presented in Section III and Section IV, respectively. Section V shows the simulation results and the corresponding explanation. The conclusion of this paper is discussed in Section VI.

\section{System Model}

\begin{figure*}[htb]
 \centering
 \includegraphics[width=0.9\linewidth]{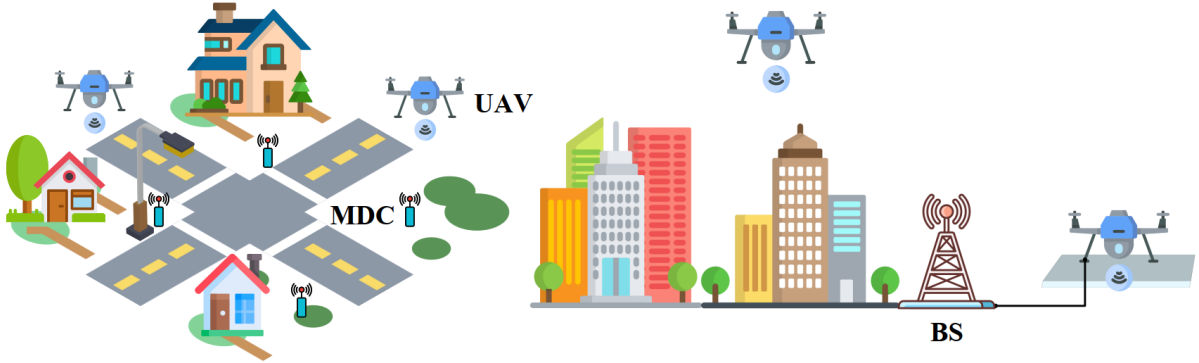}
 \caption{System model.}
 \label{fig:System_Model}
\end{figure*}

As shown in Fig. \ref{fig:System_Model}, we consider a UAV-based uplink data collection system for Metaverse service. In a given $L\times L$ area which is beyond the coverage of 5G base station, $N$ Metaverse data collectors (MDCs) are deployed to collect delay-insensitive local data, such as offline digital currency trading and weather information, which are generated by Metaverse users or the sensors \cite{DigitalCurrency2022review}, \cite{DigitalCurrency2020towards}. The location of  MDC $n$ is denoted by $(x_n,y_n,0)$. MDCs are assumed to have enough energy but limited transmission power.

To synchronize the local data with the Metaverse server, one mobile base station (MBS) carried by UAV is deployed to collect the local data saved at MDCs through $M$ channels. Each MDC can occupy only one channel, but multiple MDCs are able to share one channel. The set of MDCs in channel $m$ is denoted by $\mathcal{N}_m$, and the number of MDCs in the set is denoted as $N_m$. We assume that the UAV flies at a fixed height $H$, and the location of UAV is denoted by $(x_{\text{uav}}[t], y_{\text{uav}}[t], H)$. Once the data is received by the MBS, MDCs clear the historical data and get ready for the future data collection. In this paper, we assume that the local data size of each receiver is $U$.

\subsection{Channel Settings}
According to the experimental characterization of the vehicle-to-infrastructure radio channels in suburban environments implemented by M. Yusuf et al, the small-scale fading of the strongest path is found to be Rician distributed \cite{channel}.

The channel gain between UAV and MDC $n$ in channel $m$ and time slot $t$ is given by \cite{3DTraj}
\begin{align}
{{h}_{n,m}}[t]=\sqrt{{{\beta }_{n}}[t]}{{g}_{n,m}}[t],
\label{h}
\end{align}
where ${{\beta }_{n}}[t]$ denotes the large-scale average channel gain at time slot $t$, and ${{g}_{n,m}}[t]$ denotes the small-scale fading coefficient, which is modelled as Rician fading. ${{\beta }_{n}}[t]$ and ${{g}_{n,m}}[t]$ are given by
\begin{align}
{{\beta }_{n}}[t]={{\beta }_{0}}d_{n}^{-\alpha }[t],
\label{beta}
\end{align}
and
\begin{align}
{{g}_{n,m}}[t]=\sqrt{\frac{K}{K+1}}g+\sqrt{\frac{1}{K+1}}\tilde{g},
\label{g}
\end{align}
where $\beta_0$ denotes  the channel gain at the reference distance ${{d}_{0}}=1$m, $\alpha$ denotes the path loss exponent, which varies from 2 to 6 (in this paper we assume that $\alpha=2$). $g$ denotes the deterministic LoS channel component with $|g|=1$, which denotes the randomly scattered component. The Rician factor is denoted by $K$. $d_{n}[t]$ denotes the distance from UAV to MDC $n$ in time slot $t$, which is given by
\begin{align}
d_{n}^{{}}[t]=\sqrt{{{({{x}_{n}}-{{x}_{\text{uav}}}[t])}^{2}}+{{({{y}_{n}}-{{y}_{\text{uav}}}[t])}^{2}}+{{H}^{2}}}.
\label{d}
\end{align}
The channel-to-noise-ratio (CNR) is given by
\begin{align}
    {{\Gamma }_{n,m}}[t]=\frac{{{h}_{n,m}}[t]}{{{B\sigma }^{2}}}
\end{align}
where $\sigma^2$ denotes the power of additive white Gaussian noise (AWGN) at the receiver.
The signal to interference plus noise ratio (SINR) of MDC $n$ in channel $m$ in time slot $t$ is given by
\begin{align}
{{\gamma }_{n,m}}[t]=\frac{{{p}_{n,m}}[t]{{\Gamma }_{n,m}}[t]}{1+\sum\limits_{i=1}^{\left| {{N}_{m}} \right|-1}{{{p}_{i,m}}[t]{{\Gamma }_{i,m}}[t]}},
\end{align}
where $p_{n,m}$ denotes the transmission power of MDCs.
Thus, the transmission rate of MDC $n$ in channel $m$ and time slot $t$ is given by
\begin{align}
    {{R}_{n,m}}[t]=B{{\log }_{2}}(1+\gamma).
\end{align}

\section{Double-agent Policy Generation Model }
\begin{figure}[htb]
 \centering
 \includegraphics[width=0.9\linewidth]{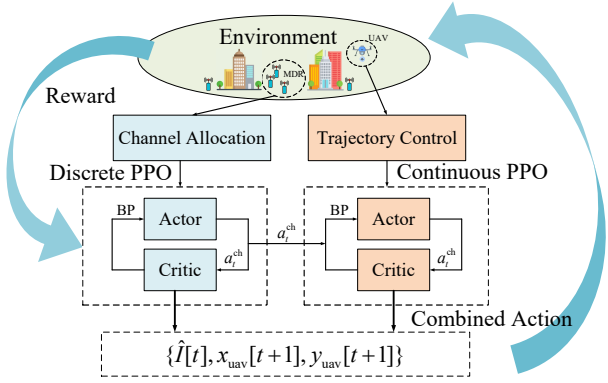}
 \caption{Double-agent policy generation model.}
 \label{solution}
\end{figure}

In this section, we introduce the double-agent policy generation model based on PPO (PPO-PPO) for channel allocation and UAV trajectory control, which is shown in Fig. \ref{solution}.

The objective is to minimize the total required time for UAV to finish collecting the data saved at MDCs with the constraint of maximum UAV speed by optimizing channel allocation indicator matrix ${\textbf{I}}[t]$, and UAV trajectory ${\{{x}_\text{uav}}[t],{{y}_\text{uav}}[t]\}$.
Each agent only focuses on a specific type of variable, and the values of other variables are loaded from the results of another agent in the previous step. In each step, the discrete proximal policy optimization (PPO) agent generates the channel allocation according to its policy, and forwards the result to the continuous PPO agent for trajectory generation. The combined action is generated by concatenating the output of two RL agents which interact with the environment to get reward for both RL agents.

\subsection{Discrete Agent for Channel Allocation}
In this subsection, we will introduce the action space, state space and reward settings of the discrete agent for channel allocation.
\subsubsection{Action of the Discrete Agent}
Intuitively, the channel allocation indicator $\textbf{I}[t]$ can be defined as an one-hot matrix, i.e., ${{I}_{n,m}}[t]\in \{0,1\}$ denotes if channel $m$ is selected by MDC $n$. An example with the number of users $N=4$ and number of channels $M=3$ is given by
\begin{align}
\textbf{I}[t]=\left[ \begin{matrix}
   {{I}_{1,1}}[t] & {{I}_{1,2}}[t] & {{I}_{1,3}}[t]  \\
   {{I}_{2,1}}[t] & {{I}_{2,2}}[t] & {{I}_{2,3}}[t]  \\
   {{I}_{3,1}}[t] & {{I}_{3,2}}[t] & {{I}_{3,3}}[t]  \\
   {{I}_{4,1}}[t] & {{I}_{4,2}}[t] & {{I}_{4,3}}[t]\\
\end{matrix} \right],
\end{align}
whose dimension is $N\times M$.
The one-hot definition of $\textbf{I}[t]$ is intuitive but increases the dimension of action space. To reduce the dimension, we re-define the channel allocation indicator matrix as
$\hat{\textbf{I}}[t]$, whose elements are ${{\hat{I}}_{n}}[t]\in \{0,1,..,M\}$. Under this definition, ${{\hat{I}}_{n}}[t]=m$ indicates that MDC $n$ is assigned with channel $m$, and ${{\hat{I}}_{n}}[t]=0$ indicates that it is not assigned with any channel.

\begin{figure}[htb]
 \centering
 \includegraphics[width=0.9\linewidth]{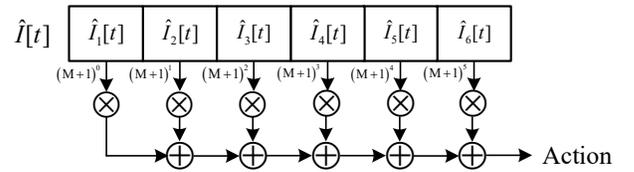}
 \caption{Action encoding.}
 \label{Encode_Action}
\end{figure}

As shown in Fig. \ref{Encode_Action}, the action of the agent is encoded according to the channel allocation indicator matrix. The encoded action is given by
\begin{align}
a_{t}^{\text{ch}}=\sum\limits_{n=1}^{N}{{{{\hat{I}}}_{n}}[t]{{(M+1)}^{n-1}}}
\end{align}

\subsubsection{State of the Discrete Agent}
The decisions of RL agents are generated based on the current state. In this paper, the state of the discrete agent includes the channel gain and the remaining data at MDCs in the current step. The state of the discrete agent is concatenated by two parts, which is given by

\begin{align}
S_{t}^{\text{ch}}=\{{{U}_{\text{res}}}[t],h[t]\},
\end{align}
where ${U}_{\text{res}}$ denotes the matrix of remaining data in MDCs, and $h[t]$ denotes the matrix of channel gain at $t^{th}$ step.

\subsubsection{Reward of the Discrete Agent}
The optimization objective in this paper is the required time for UAV to finish the data collection mission, i.e., to minimize the number of steps in each episode. Intuitively, the more steps the agent takes, the less reward it should receive.
Thus, we set a time-based penalty $r_{t}^{\text{time}}$ with negative value in each step to build the connection between reward and our objective.
If the agent fails to finish the mission in given time limit $T_{\text{max}}$, it will receive a failure penalty $r_{t}^{\text{fail}}$.

The time-based penalty $r_{t}^{\text{time}}$ is further modified according to the data size collected by UAV in the current step to give higher reward to the actions which result to larger transmission rate. The reward of the discrete agent is given by

\begin{equation}
r_{t}^{\text{ch}}=\left\{ \begin{aligned}
  & \frac{r_{t}^{\text{time}}}{U}\sum\limits_{n=1}^{N}{\sum\limits_{m=1}^{M}{{{t}_{\text{slot}}}{{R}_{n,m}}[t],\text{ }if\text{ }t\le {{T}_{\max }}}} \\
 & r_{t}^{\text{fail}},\text{ }if\text{ }t>{{T}_{\max }}
\end{aligned} \right.
\end{equation}

\subsection{Continuous Agent for Trajectory Optimization}
The trajectory of UAV is optimized by a continuous RL agent, whose action, state and reward are defined as follows.

\subsubsection{Action of the continuous agent}

\begin{figure}[htb]
 \centering
 \includegraphics[width=0.9\linewidth]{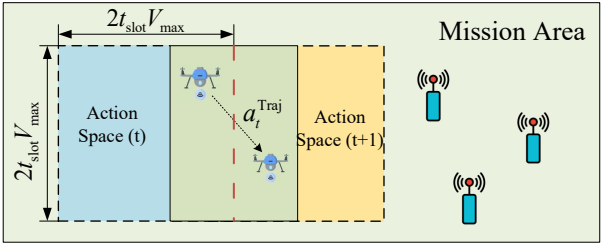}
 \caption{Action space of the continuous agent.}
 \label{Traj_space}
\end{figure}

As shown in Fig. \ref{Traj_space}, the action of the continuous agent $a_t^{\text{traj}}$ determines the location of UAV in the next step. $a_{t}^{\text{traj}}$ is defined as

\begin{align}
a_{t}^{\text{traj}}=\{a_{t}^{x},a_{t}^{y}\},a_{t}^{x},a_{t}^{y}\in \left[ -{{t}_{\text{slot}}}{{V}_{\max }},{{t}_{\text{slot}}}{{V}_{\max }} \right],
\end{align}
where $a_{t}^{x},a_{t}^{y}$ denote the movement of UAV on the x-axis and y-axis respectively.

\subsubsection{State of the Continuous Agent}
The state of the continuous agent is similar to the discrete agent, which includes the current channel gain $h[t]$ and remaining data at MDCs ${{U}_{\text{res}}}[t]$. In addition, the current horizontal location of UAV $\left( {{x}_{\text{uav}}}[t],{{y}_{\text{uav}}}[t] \right)$ is also included in the state $S_{t}^{\text{traj}}$, which is given by

\begin{align}
   S_{t}^{\text{traj}}=\left\{ {{U}_{\text{res}}}[t],h[t],\left( {{x}_{\text{uav}}}[t],{{y}_{\text{uav}}}[t] \right) \right\}
\end{align}

\subsubsection{Reward of the Continuous Agent}
The reward of the continuous agent is modified based on $r_{t}^{\text{ch}}$. We give additional penalty to the agent if the location of UAV exceeds reasonable region to regularize the trajectory decision. The reward of the continuous agent is given by

\begin{equation}
r_{t}^{\text{traj}}=\left\{ \begin{aligned}
  & r_{t}^{\text{ch}},\text{ }if\text{ }{{x}_{\text{uav}}}[t]\in [{{x}_{\min }},{{x}_{\max }}],{{y}_{\text{uav}}}[t]\in [{{y}_{\min }},{{y}_{\max }}] \\
 & r_{t}^{\text{ch}} + r_{t}^{\text{penalty}},\text{ } otherwise
\end{aligned} \right.
\end{equation}

\section{Implementation of Proximal Policy Optimization (PPO)}
PPO is a state-of-art on policy reinforcement learning algorithm which supports both discrete and continuous actions spaces. In this section, we introduce the preliminary and implementation of PPO algorithm for discrete agent (channel allocation) and continuous agent (UAV trajectory optimization).

\subsection{Implementation of Continuous and Discrete PPO}

\subsubsection{Critic Network}

The critic network is responsible to give scores to the actor according to the current state. The architectures of both discrete and continuous critic networks are the same, which consists of multiple fully connected layers.

\textbf{Loss function of continuous and discrete critic networks} are given by
\begin{align}
J^{\text{traj}}(\phi )={{\left[ V_{\phi }^{\text{traj}}(s_{t}^{\text{traj}})-\left( r_{t}^{\text{traj}}+\gamma V_{{{\phi }'}}^{\text{traj}}(s_{t+1}^{\text{traj}}) \right) \right]}^{2}},
\label{a_loss_con}
\end{align}

\begin{align}
J^{\text{ch}}(\phi )={{\left[ V_{\phi }^{\text{ch}}(s_{t}^{\text{ch}})-\left( r_{t}^{\text{ch}}+\gamma V_{{{\phi }'}}^{\text{ch}}(s_{t+1}^{\text{ch}}) \right) \right]}^{2}},
\label{a_loss_dis}
\end{align}
where $L_{t}^{\text{traj}}(\phi )$ and $L_{t}^{\text{ch}}(\phi )$ denote the loss function for the critic network of continuous and discrete agent respectively.   $V_{{{\phi }'}}^{\text{traj}}(s_{t+1}^{\text{traj}})$ and   $V_{{{\phi }'}}^{\text{ch}}(s_{t+1}^{\text{ch}})$ are the state value estimations generated by the old critic networks $\phi _{\text{traj}}^{'}$ and $\phi_{\text{ch}}^{'}$ respectively, which are saved in during the interaction with environment. $V_{\phi }^{\text{traj}}(s_{t}^{\text{traj}})$ and $V_{\phi }^{\text{ch}}(s_{t}^{\text{ch}})$ are the state value estimations generated by the current critic networks $\phi _{\text{traj}}^{{}}$ and $\phi _{\text{ch}}^{{}}$ , which are updated in each training iteration.

\subsubsection{Actor Network}
\begin{figure}[htb]
 \centering
 \includegraphics[width=0.9\linewidth]{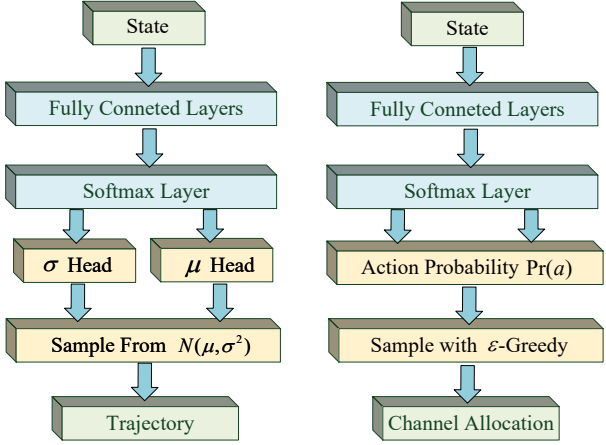}
 \caption{Actor Network Architecture.}
 \label{Actor}
\end{figure}

As shown in Fig. \ref{Actor}, the architecture of discrete and continuous actor network are different due to the difference in action space.

\textbf{The continuous actor network} for trajectory control is a network for value approximation, which outputs a $\mu$ head and a $\sigma$ head which denotes the mean and variance of Gaussian distributions respectively. Each head includes two variables, i.e., $\{{{\mu }_{x}},{{\mu }_{y}}\}$ and  $\{{{\sigma }_{x}},{{\sigma }_{y}}\}$, which denotes the $x$-axis and $y$-axis respectively. The action $\{{{a}_{x}}[t],{{a}_{y}}[t]\}$ is generated by sampling from the obtained distribution $\mathcal {N} ({{\mu }_{x}},\sigma _{x}^{2})$ and $\mathcal{N}({{\mu }_{y}},\sigma _{y}^{2})$.

\textbf{The discrete actor network} for channel allocation is a network for classification, which outputs the probabilities $\text{Pr}(a)$ of each action. The agent sample its action from the obtained action probabilities with $\varepsilon$-greedy, i.e., the output action is generated by sampling from $\text{Pr}(a)$ with probability $1-\epsilon$, and selected randomly with probability $\epsilon$ for exploration. The output action of the discrete actor network is encoded, which will be decoded into one-hot indicators before being utilized for further calculation.

\textbf{Loss functions of actor networks} in our implementation adopt the trick of clipping to simplify the calculation, which is proposed by J. Schulman et al \cite{PPO_origin}.

The PPO-PPO algorithm is summarized in \textbf{Algorithm 1}.

\begin{figure}[!t]
        \renewcommand{\algorithmicrequire}{\textbf{Initiate:}}
        \renewcommand{\algorithmicensure}{\textbf{Output:}}
        \begin{algorithm}[H]
            \caption{\label{alg:HCORL} PPO-PPO}
            \begin{algorithmic}[1]
                \REQUIRE Remaining data at MDCs, UAV location, network parameters of discrete and continuous agent
                \FOR{iteration $t$ = $1,2,..$}
                    \STATE Discrete agent execute action according to the current state and policy $\pi _{{{\theta }'}}^{\text{ch}}\left( a_{t}^{\text{ch}}|s_{t}^{\text{ch}} \right)$ to obtain the channel allocation indicator matrix $\hat{\textbf{I}}[t]$
                    \STATE With given $\hat{\textbf{I}}[t]$, the continuous agent for trajectory control execute action according to the current state and policy $\pi _{\theta' }^{\text{traj}}\left( a_{t}^{\text{traj}}|s_{t}^{\text{traj}} \right)$ $\hat{I}[t]$
                    \STATE Agent interact with environment to get reward $r_{t}^{\text{ch}}$ and $r_{t}^{\text{traj}}$ for discrete agent and continuous agent respectively
                    \STATE Update state $s_{t}^{\text{traj}}\leftarrow s_{t+1}^{\text{traj}}$, $s_{t}^{\text{ch}}\leftarrow s_{t+1}^{\text{ch}}$
                    \STATE Save trajectory $\left( s_{t}^{\text{ch}},a_{t}^{\text{ch}},r_{t}^{\text{ch}},s_{t+1}^{\text{ch}},V_{{{\phi }'}}^{\text{ch}}(s_{t}^{\text{ch}}) \right)$ and
                    $\left( s_{t}^{\text{traj}},a_{t}^{\text{traj}},r_{t}^{\text{traj}},s_{t}^{\text{traj}},V_{{{\phi }'}}^{\text{traj}}(s_{t}^{\text{traj}}) \right)$
                    \FOR{every $i$ iterations}
                        \STATE Shuffle data order and make batch with size $bs$.
                        \FOR{$j$=$0,1,...,\frac{T}{bs}-1$}
                            \STATE Calculate loss functions of critic and actor networks and update network parameters by gradient ascent
                        \ENDFOR
                    \ENDFOR
                \ENDFOR
            \end{algorithmic}
        \end{algorithm}
\label{alg_HCORL}
\vspace{-0.5cm}
\end{figure}

\section{Simulation Results}
The performance of our proposed double-agent reinforcement learning approach for Metaverse data collecting is tested and compared with two benchmark scenarios (DQN-PPO and duelling DQN-PPO), whose discrete agents are replaced with DQN or duelling DQN algorithm respectively. The simulation settings are given in Table \ref{table:parameter}.

\begin{table}[tbp]
\caption{Constant Parameter Setting} \label{table:parameter}
\begin{center}
\begin{tabular}{c c}
\toprule[1pt]
 Parameter and Physical Meaning            & Value                  \\ \hline
  Number of channels($M$)                  & $3$                    \\
  Default number of users ($N$)            & $5$                    \\
  Bandwidth ($B$)                          & $5$MHz                 \\
  Transmission power of MDCs               & $5$W                   \\
  Frequency ($f$)                          & $28$GHz (5G spectrum)  \\
  Power of Gaussian noise ($\sigma^2$)     & $5\times {{10}^{-8}}$W \\
  Maximum speed of UAV                     & $10$m/s                \\
  Mission area size ($L$)                  & $200$m                 \\
\bottomrule[1pt]
\end{tabular}
\end{center}
\end{table}

\begin{figure}[htb]
 \centering
 \includegraphics[width=0.9\linewidth]{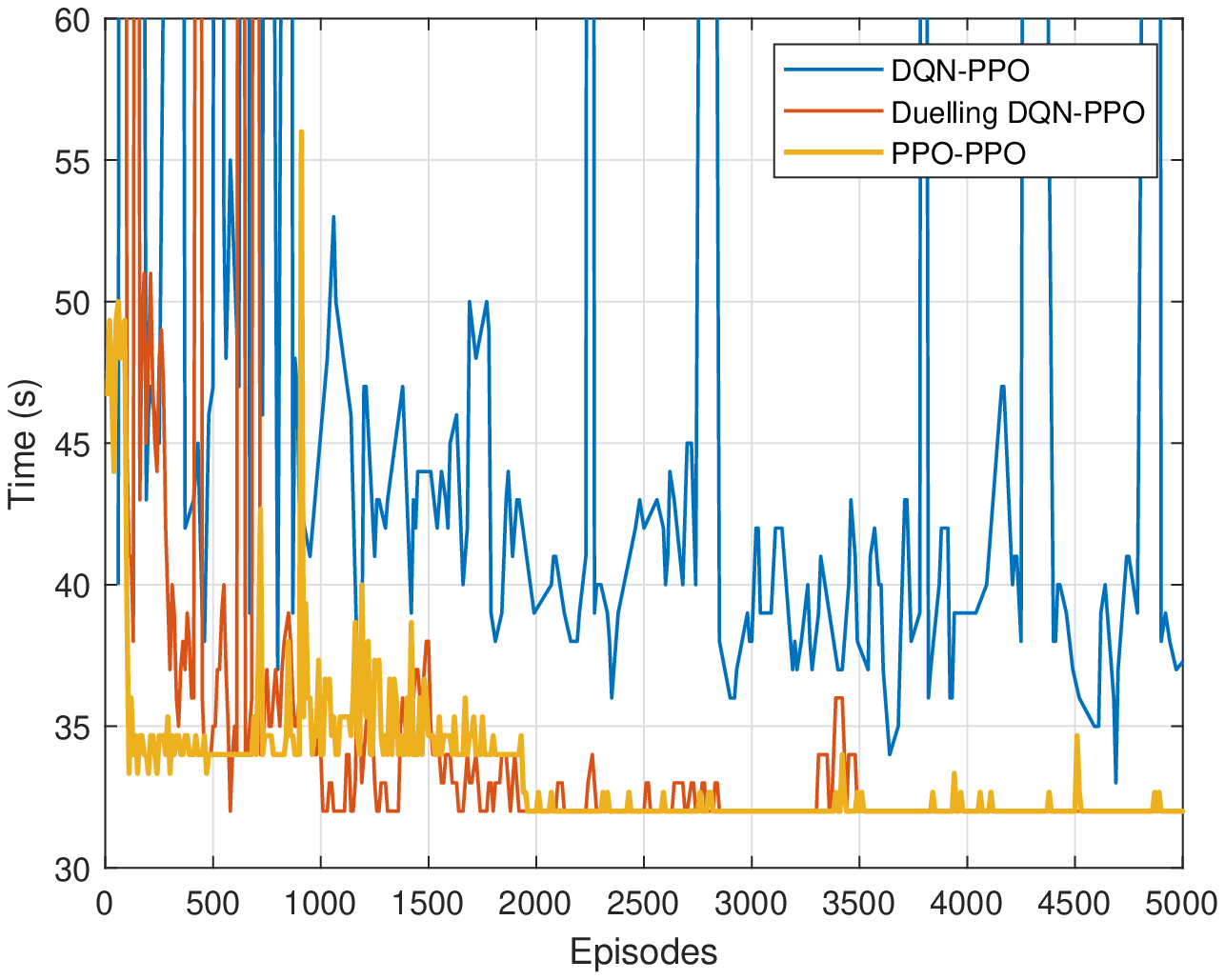}
 \caption{Comparison of required time to finish mission with data size $U=50$Mb.}
 \label{fig:MissionTime_50M}
\end{figure}

Fig. \ref{fig:MissionTime_50M} presents the required time to complete data collecting mission of our proposed algorithm and two benchmark algorithms with given data size $U=50$Mb. At the beginning of the training process (0-1000 episodes), all three algorithms are unstable because the reasonable policy has not been established, and the agents are exploring the environment frequently. From 1000 episodes to 2000 episodes, our proposed PPO-PPO algorithm shows the tendency of convergence while the benchmark DQN-PPO algorithm is still very unstable. The duelling DQN-PPO algorithm also starts to finish the mission within a shorter time period, but is less stable than the PPO-PPO algorithm. DQN-PPO algorithm shows poor convergence performance in this task, but both PPO-PPO and duelling DQN-PPO algorithms are able to converge within 5000 episodes with similar performance due to their common implementation of the advantage function.

\begin{figure}[htb]
 \centering
 \includegraphics[width=0.9\linewidth]{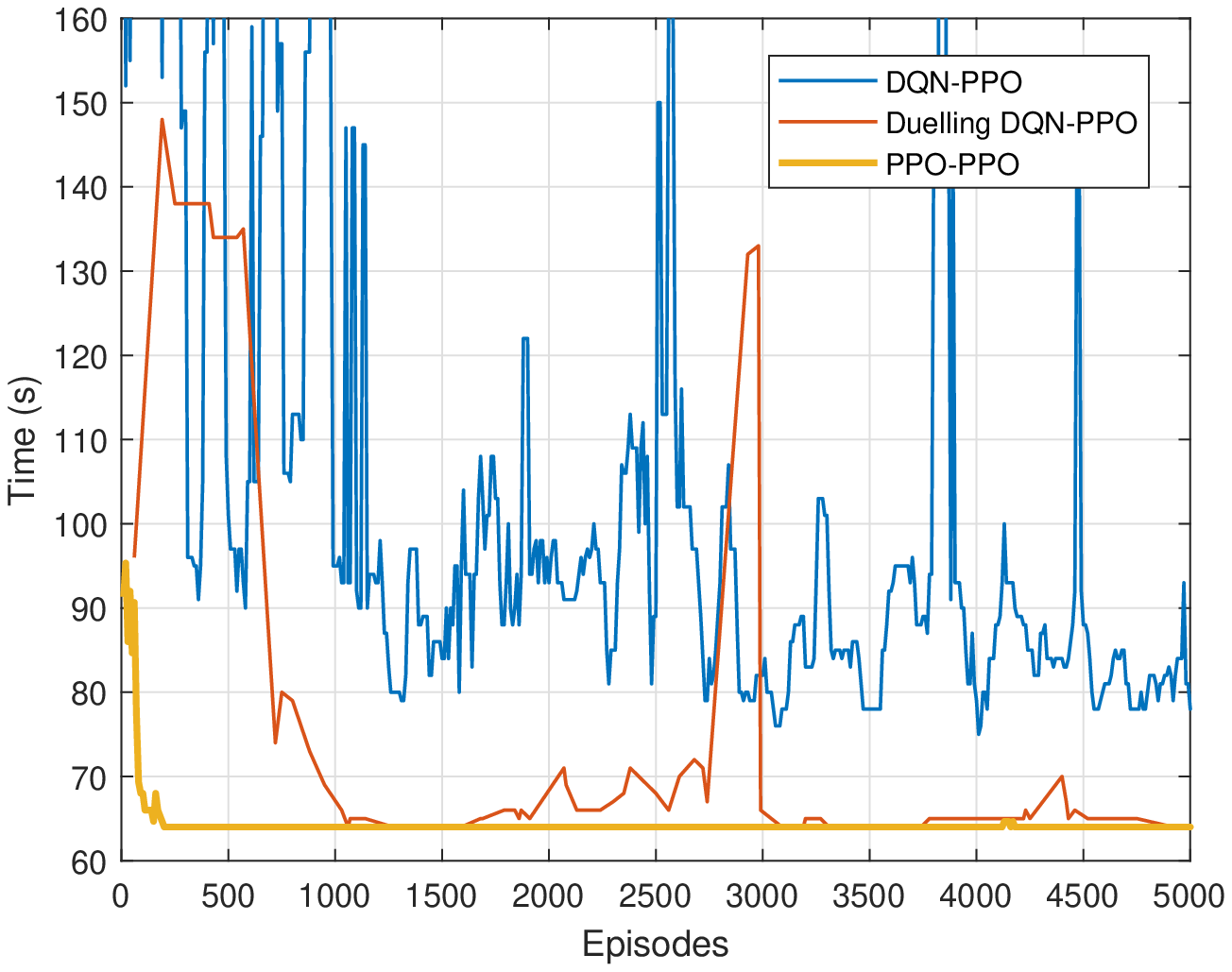}
 \caption{Comparison of required time to finish mission with data size $U=100$Mb.}
 \label{fig:MissionTime_100M}
\end{figure}

Fig. \ref{fig:MissionTime_100M} presents the mission completing time experiment with a similar parameter setting as in Fig. \ref{fig:MissionTime_50M}, but the data size is increased to $U=100$Mb. All three algorithms need more time to finish the data collecting mission due to larger data size, and the PPO-PPO algorithm shows similar convergence performance as in Fig. \ref{fig:MissionTime_50M}. However, the dueling DQN-PPO algorithm becomes unstable in the training process, i.e., some sudden increase in the required time. The superior stability of PPO over dueling DQN can be attributed to its policy update constraint by equipping it with a KL-divergence penalty between the old policy (the policy for sampling data) and the updated policy (the policy used for training and evaluating).

\begin{figure}[htb]
 \centering
 \includegraphics[width=0.9\linewidth]{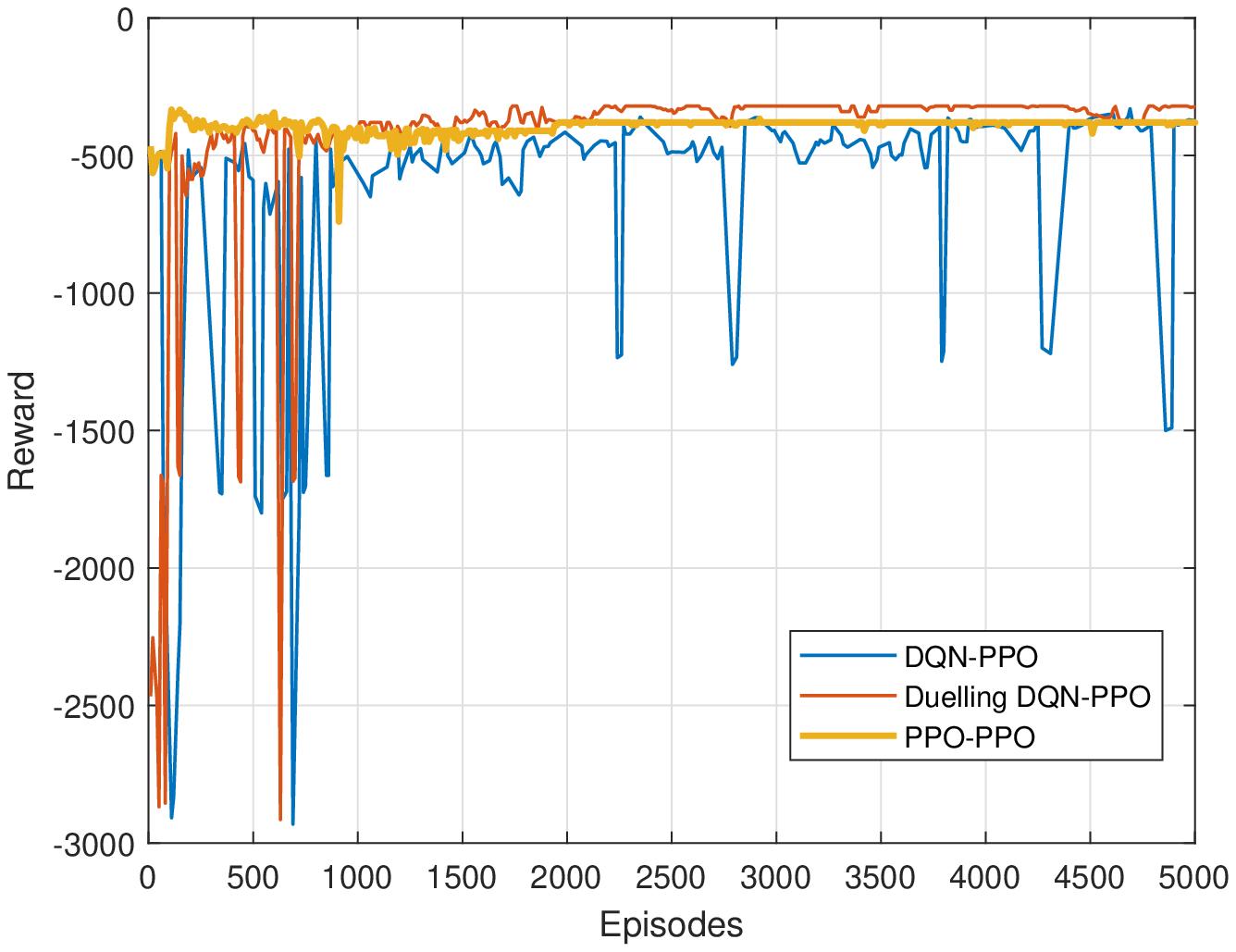}
 \caption{Comparison of reward with data size $U=50$Mb.}
 \label{fig:Reward_50M}
\end{figure}

\begin{figure}[htb]
 \centering
 \includegraphics[width=0.9\linewidth]{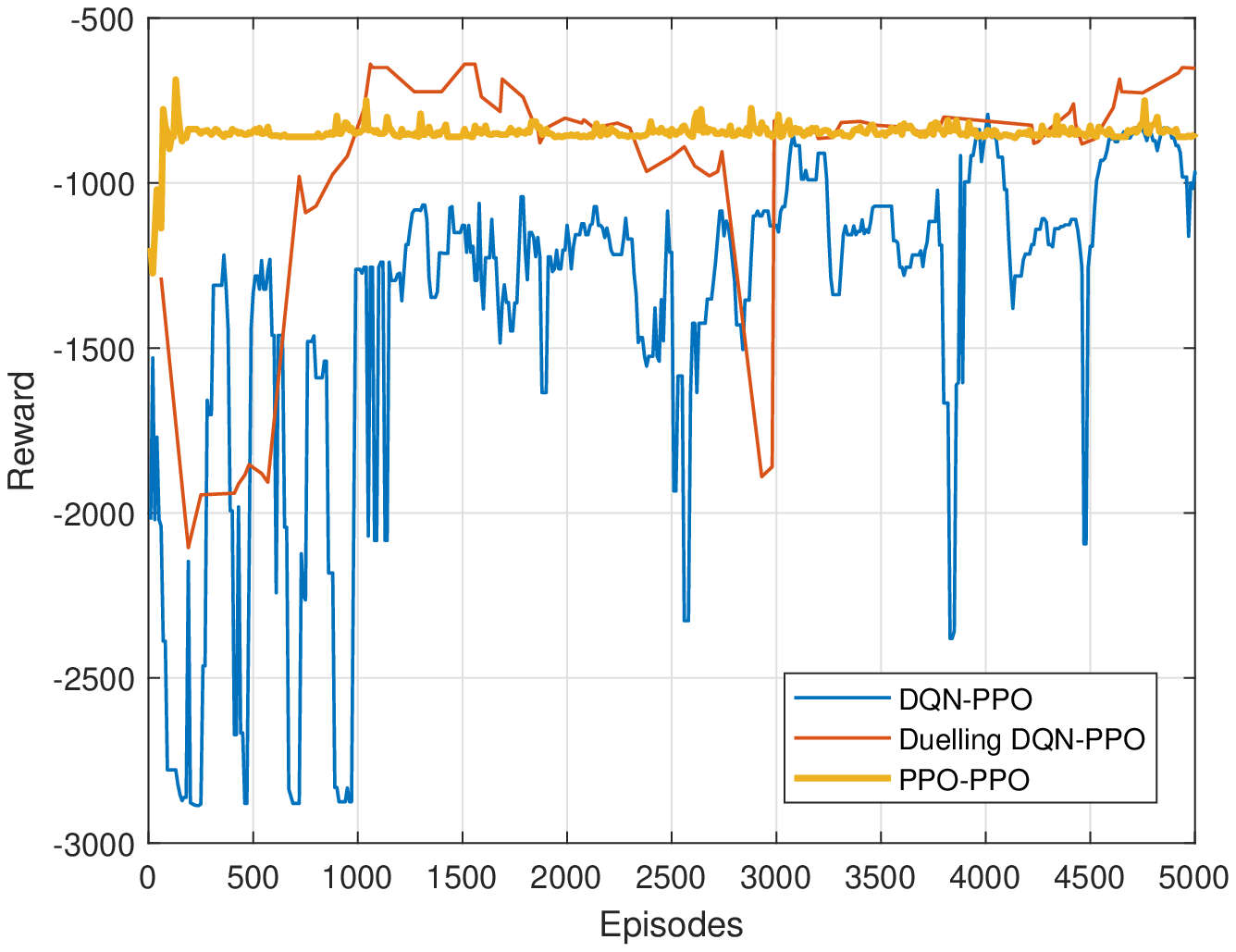}
 \caption{Comparison of reward with data size $U=100$Mb.}
 \label{fig:Reward_100M}
\end{figure}

Fig. \ref{fig:Reward_50M} and Fig. \ref{fig:Reward_100M} are the corresponding rewards in the training processes of Fig. \ref{fig:MissionTime_50M} and Fig. \ref{fig:MissionTime_100M} respectively. We consider the reward given to the agent as guidance but not the exact objective function in the implementation of reinforcement learning algorithm. The tendencies of the reward and the required time are highly similar although they are generated from different formulas, which indicates that our reward design successfully leads the agent to learn a better policy.

\begin{figure}[htb]
 \centering
 \includegraphics[width=0.9\linewidth]{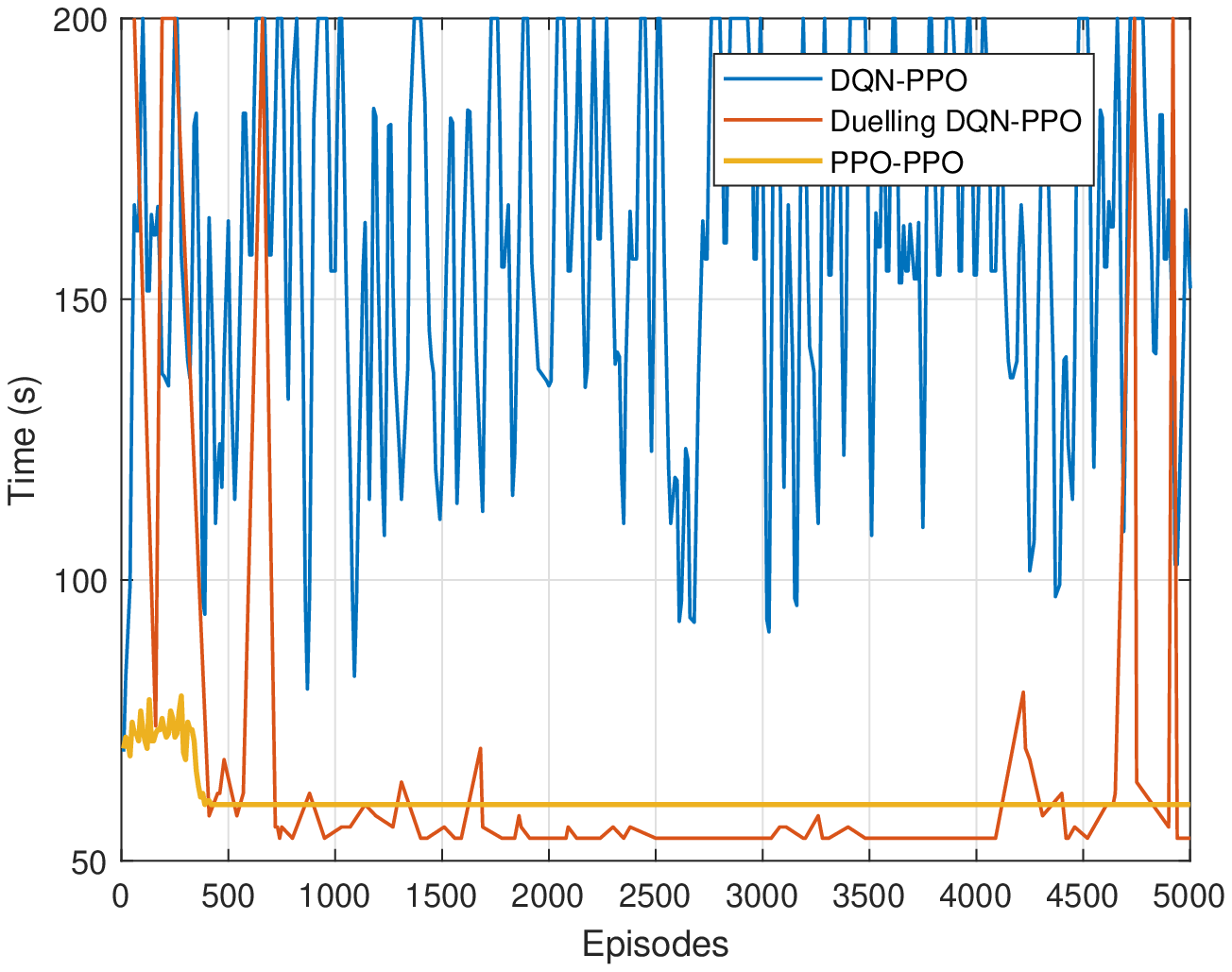}
 \caption{Comparison of reward with data size $U=50$Mb and $8$ users.}
 \label{fig:MissionTime_50M_8user}
\end{figure}

The mission completing time comparison for the case with eight users is shown in Fig. \ref{fig:MissionTime_50M_8user}. The duelling DQN-PPO algorithm shows similar average performance as the PPO-PPO algorithm but less stability, i.e., the required time sometimes jumps to extremely large values. Taking the stability into consideration, the PPO-PPO algorithm is better than duelling DQN-PPO algorithm in general. The DQN-PPO algorithm is obviously not able to converge in this experiment, so we do not consider it a candidate for our double-agent reinforcement learning algorithm.

\section{Conclusion}
In this paper, we propose a double-agent reinforcement architecture for data collecting and synchronization in Metavese, and adopt PPO algorithm for both discrete and continuous agents. Two agents with different action space and state space work in a cascade manner for channel allocation and UAV trajectory control to form a combined action in each iteration. Our experiments indicate the advantage of the PPO-PPO in both the required time for the mission and the stability. In future work, we will consider transmission power allocation and test the performance of other state-of-art reinforcement learning algorithms in our proposed architecture.


\begin{thebibliography}{99}
\bibitem{wang2022mobile}
Y. Wang and J. Zhao, ``Mobile Edge Computing, Metaverse, 6G Wireless Communications, Artificial Intelligence, and Blockchain: Survey and Their Convergence,” arXiv preprint arXiv:2209.14147, 2022.

\bibitem{AllYouNeedToKnow}
L.-H. Lee, T. Braud, P. Zhou, L. Wang, D. Xu, Z. Lin, A. Kumar, C. Bermejo, and P. Hui, ``All one needs to know about metaverse: A complete survey on technological singularity, virtual ecosystem, and research agenda,” arXiv preprint arXiv:2110.05352, 2021.

\bibitem{GlobeCom2022}
P. Si, J. Zhao, H. Han, K.-Y. Lam, and Y. Liu, “Resource Allocation and Resolution Control in the Metaverse with Mobile Augmented Reality,” arXiv preprint arXiv:2209.13871, 2022.

\bibitem{Terence2022resource}
T. J. Chua, W. Yu, and J. Zhao, ``Resource allocation for mobile metaverse with the Internet of Vehicles over 6G wireless communications: A deep reinforcement learning approach,” arXiv preprint arXiv:2209.13425, 2022.

\bibitem{MetaEdu2020}
H.-C. Han et al., ``From visual culture in the immersive metaverse to visual cognition in education,” in Cognitive and Affective Perspectives on Immersive Technology in Education. IGI Global, 2020, pp. 67–84.

\bibitem{Exploration of educational possibilities by four metaverse type}
J.-E. Yu, ``Exploration of educational possibilities by four metaverse types in physical education,” Technologies, vol. 10, no. 5, p. 104, 2022.

\bibitem{MetaTrad_1}
Murat Yilmaz, Tuna Hacalo{\u{g}}lu and Paul Clarke , ``Examining the use of non-fungible tokens (NFTs) as a trading mechanism for the metaverse,” in
European Conference on Software Process Improvement. Springer, 2022, pp. 18–28.

\bibitem{MetaGame}
A. Jungherr and D. B. Schlarb, ``The Extended Reach of Game Engine Companies: How Companies Like Epic Games and Unity Technologies Provide Platforms for Extended Reality Applications and the Metaverse,” Social Media+ Society, vol. 8, no. 2, 2022.

\bibitem{5GSurvey}
M. Agiwal, A. Roy, and N. Saxena, ``Next generation 5G wireless networks: A comprehensive survey,” IEEE Communications Surveys \& Tutorials, vol. 18, no. 3, pp. 1617–1655, 2016.

\bibitem{NFC}
J. Refonaa, G. G. Sebastian, D. Ramanan, and M. Lakshmi, ``Effective identification of black money and fake currency using NFC, IoT and android,” in 2018 International Conference on Communication, Computing and Internet of Things (IC3IoT). IEEE, 2018, pp. 275–278.

\bibitem{OfflineTrade}
J. Kohli and A. Kohli, ``A Study of Preference for Online Trading Account versus Offline Trading Account by Different Age Group,” From the Desk of Editor, p. 42, 2020.

\bibitem{sensorWild}
P. V. Mane-Deshmukh, ``Designing of Wireless Sensor Network to Protect Agricultural Farm from Wild Animals,” i-Manager’s Journal on Information Technology, vol. 7, no. 4, p. 30, 2018.

\bibitem{RuiZhangUAV}
Q. Wu, Y. Zeng, and R. Zhang, ``Joint trajectory and communication design for multi-UAV enabled wireless networks,” IEEE Transactions on Wireless Communications, vol. 17, no. 3, pp. 2109–2121, 2018.

\bibitem{SiIoTJ}
W. Lu, P. Si, Y. Gao, H. Han, Z. Liu, Y. Wu, and Y. Gong, ``Trajectory and Resource Optimization in OFDM-Based UAV-Powered IoT Network,” IEEE Transactions on Green Communications and Networking, vol. 5, no. 3, pp. 1259–1270, 2021.

\bibitem{UAVIoV}
J. S. Ng, W. Y. B. Lim, H.-N. Dai, Z. Xiong, J. Huang, D. Niyato, X.-S. Hua, C. Leung, and C. Miao, ``Joint auction-coalition formation framework for communication-efficient federated learning in UAV-enabled internet of vehicles,” IEEE Transactions on Intelligent Transportation Systems, vol. 22, no. 4, pp. 2326–2344, 2020.

\bibitem{Cui_MultiUAV}
J. Cui, Y. Liu, and A. Nallanathan, ``Multi-agent reinforcement learning-based resource allocation for UAV networks,” IEEE Transactions on Wireless Communications, vol. 19, no. 2, pp. 729–743, 2019.

\bibitem{luong2021deep}
P. Luong, F. Gagnon, L.-N. Tran, and F. Labeau, ``Deep reinforcement learning-based resource allocation in cooperative UAV-assisted wireless networks,” IEEE Transactions on Wireless Communications, vol. 20, no. 11, pp. 7610–7625, 2021.

\bibitem{UAVland}
A. Rodriguez-Ramos, C. Sampedro, H. Bavle, P. De La Puente, and P. Campoy, ``A deep reinforcement learning strategy for UAV autonomous landing on a moving platform,” Journal of Intelligent \& Robotic Systems, vol. 93, no. 1, pp. 351–366, 2019.

\bibitem{Hybrid_MultiAgent_1}
H. Fu, H. Tang, J. Hao, Z. Lei, Y. Chen, and C. Fan, ``Deep multi-agent reinforcement learning with discrete-continuous hybrid action spaces,”
arXiv preprint arXiv:1903.04959, 2019.

\bibitem{Hybrid_multiAgent_2}
N. Jiang, Y. Deng, and A. Nallanathan, ``Deep reinforcement learning for discrete and continuous massive access control optimization,” in ICC
2020-2020 IEEE International Conference on Communications (ICC). IEEE, 2020, pp. 1–7.

\bibitem{DigitalCurrency2022review}
Y. Chu, J. Lee, S. Kim, H. Kim, Y. Yoon, and H. Chung, ``Review of Offline Payment Function of CBDC Considering Security Requirements,” Applied Sciences, vol. 12, no. 9, p. 4488, 2022.

\bibitem{DigitalCurrency2020towards}
M. Christodorescu, W. C. Gu, R. Kumaresan, M. Minaei, M. Ozdayi, B. Price, S. Raghuraman, M. Saad, C. Sheffield, M. Xu et al., ``Towards a two-tier hierarchical infrastructure: An offline payment system for Central Bank digital currencies,” arXiv preprint arXiv:2012.08003, 2020.

\bibitem{channel}
M. Yusuf, E. Tanghe, F. Challita et al., ``Experimental characterization of V2I radio channel in a suburban environment,” in 2019 13th European Conference on Antennas and Propagation (EuCAP). IEEE, 2019, pp. 1–5.

\bibitem{3DTraj}
C. You and R. Zhang, ``3D trajectory optimization in Rician fading for UAV-enabled data harvesting,” IEEE Transactions on Wireless Communications, vol. 18, no. 6, pp. 3192–3207, 2019.

\bibitem{PPO_origin}
J. Schulman, F. Wolski, P. Dhariwal, A. Radford, and O. Klimov, ``Proximal policy optimization algorithms,” arXiv preprint arXiv:1707.06347, 2017.

\end{thebibliography}
\end{document}